\documentclass[conference, 10pt]{IEEEtran}

\usepackage[switch]{lineno}
\usepackage{cite}
\usepackage{amsmath,amssymb,amsfonts}
\usepackage{mathtools}
\usepackage[algo2e]{algorithm2e}
\usepackage[noend]{algpseudocode}
\usepackage{textcomp}
\usepackage{subfigure}
\usepackage{array}
\usepackage{makecell}
\usepackage{hyphenat}
\usepackage{fancyhdr}
\usepackage{listings}
\usepackage{color}
\usepackage[dvipsnames,table,xcdraw]{xcolor}
\usepackage{graphicx}
\usepackage{booktabs}
\usepackage{algorithm}
\usepackage{algorithmicx}
\usepackage[]{algpseudocode}
\usepackage{multirow}
\usepackage{balance}
\usepackage{float}
\usepackage{pifont}
\usepackage{url}
\usepackage[dvipsnames,table,xcdraw]{xcolor}
\usepackage{comment}
\usepackage{subfigure}
\usepackage{pifont}

\usepackage{flushend}
\usepackage{hyperref}   

\algdef{SE}[SUBALG]{Indent}{EndIndent}{}{\algorithmicend\ }%
\algtext*{Indent}
\algtext*{EndIndent}

\newcommand{\MySection}[1]{
  \section{#1}
}
\newcommand{\MySubsection}[1]{
  \subsection{#1}
}

\newcommand{\MyCaption}[1]{
  \vspace{-1.0ex}
  \caption{#1}
  \vspace{-1.0ex}
}

\begin{document}
\title{
Cross-layer Visualization and Profiling of Network and I/O Communication for HPC Clusters
}

\author{\IEEEauthorblockN{
        Pouya Kousha,
        Quentin Anthony,
        Hari Subramoni and
        Dhabaleswar K. Panda
    }
    
    \IEEEauthorblockA{
        Department of Computer Science and Engineering \\
        The Ohio State University \\
        \{kousha.2, anthony.301\}@osu.edu, \{subramon, panda\}@cse.ohio-state.edu
    }
    \vspace{-5.0ex}
}

\newcommand{\name}{INAM\xspace}
\newcommand{\cmark}{\ding{51}}%
\newcommand{\xmark}{\ding{55}}%
    
\maketitle
\begin{abstract}

Understanding and visualizing the full-stack performance trade-offs and interplay
between HPC applications, MPI libraries, the communication
fabric, and the file system is a challenging
endeavor. Designing a holistic profiling and visualization method for
HPC communication networks is challenging since different levels of communication coexist and
interact with each other on the communication fabric. A breakdown of traffic is essential to understand the interplay of different layers along with the application's communication behavior without losing a general view of network traffic.

Unfortunately, existing profiling tools are disjoint and either focus on only profiling and visualizing a few levels of the HPC stack, which limits the insights they can provide, or they provide extremely detailed information which necessitates a steep learning curve to understand. We target our profiling tool visualization to provide holistic and real-time insights into HPC communication stacks.

In this paper, we propose and implement our
visualization methods to enable holistic insight for representing the cross-stack metrics. Moreover, we propose and implement a low-overhead I/O profiling inside the communication library, collect and store the profiling information, and then study the correlation and evaluation of I/O traffic with MPI communication using a cross-stack approach by INAM. Through experimental evaluations and use cases, we demonstrate novel benefits of our cross-stack communication analysis in real-time to detect bottlenecks and understand communication performance. 

\end{abstract}

\begin{IEEEkeywords}
MPI, I/O, Visualization, Profiling, Communication
\end{IEEEkeywords}

\MySection{Introduction and Motivation}
\label{sec:intro}

Recent advances in High Performance Computing (HPC) have
provided a fast processing engine for HPC applications. HPC
systems must be highly optimized to meet the growing needs of modern HPC applications. In order 
to deliver the required computing power for the applications, large-scale HPC
clusters with Mellanox InfiniBand and
High Speed Ethernet are being deployed and used by various HPC users like 
domain scientists, HPC software developers, and HPC administrators. The Message Passing Interface (MPI)~\cite{mpi3} is a very popular parallel programming model for developing parallel scientific applications to run on HPC systems.

With the recent advances in interconnect technology and increasing cost of communication, 
providing efficient data movement between nodes on the communication fabric is essential to guarantee optimized
end-to-end solutions. In such an
ecosystem, understanding the interaction between individual nodes and switches with respect to the edge and root switches
has become even more challenging. For HPC users, tuning one job on a small node allocation is just the start. When running on a larger system concurrently running hundreds of high-performance jobs, it becomes challenging to observe the communication behaviour of individual jobs as well as interplay among jobs. Therefore, detailed and easy-to-interpret insight of the network and communication levels is needed
for all types of HPC users to identify and alleviate performance bottlenecks.

As applications are moving towards larger datasets, developers are opting to use the HPC file system to store and access the data or for check-pointing operations during the job run-time. Hence, the I/O file system plays an important role in application communication performance. Almost all HPC clusters have some set of storage nodes to handle both massive amounts of data and the compute nodes accessing these storage nodes. Therefore, HPC application run-times are becoming more I/O sensitive.

Characterizing the communication performance in a ``holistic'' manner is 
complicated and time-consuming as it depends on the interplay between 
different levels of the HPC ecosystem. The problem of identifying performance issues 
with these HPC applications is an overwhelming undertaking
---akin to finding ``a needle in a haystack''. Important components at play here include: 
1) HPC/DL applications, 2) High-Performance MPI Communication Runtime, 
3) Network Fabric, 4) I/O Subsystem, and 5) Job Scheduler. For this reason, it 
becomes essential and challenging to have novel methods to perform 
introspection, evaluation, and visualization of communication and I/O interactions 
occurring between various components of the HPC ecosystem. 
This is important since most of the application developers---in 
particular scientific researchers---do not have the skills and knowledge 
to carry out postmortem analysis of the performance 
of their applications. Thus, identifying 
the source of performance degradation for application developers might 
translate to days of saved effort.

One of the major challenges is visualizing the complicated network traffic of HPC systems. The communication traffic can unfold into different dimensions. The InfiniBand low-level communication traffic is divided into unicast and multicast communication traffic. efficient multicast patterns benefit faster communication operations~\cite{ib-multicast} since technologies like SHARP~\cite{sharp} focus on using multicast communication on switches. Running on top of InfiniBand, there is MPI traffic that could coexist with I/O traffic from other jobs. Moreover, there is received and sent data for each traffic type. Therefore, it is challenging to visualize the communication traffic breakdown while including each communication dimension. Furthermore, There is a need for interactive and straightforward visualizations to assist in better understanding the communication characterization.

Several tools exist (as described in
Table~\ref{tab:related_tools}) that have partially addressed the 
challenge of introspecting cross-stack communication in the HPC ecosystem.
However, there exists no single tool that can
holistically introspect and evaluate the communication
and I/O interactions occurring between various components while correlating
the information obtained from different sources across the HPC
ecosystem. To the best of our knowledge, there is no holistic visualization capable of showing the breakdown of the communication fabric while displaying node connectivity among the common HPC communication profiling tools mentioned in Table~\ref{tab:related_tools}.  


\begin{table}[htbp]
\centering
\MyCaption{Summary of existing communication profiling tools and their capabilities based on the interaction of the MPI runtime with application, network fabric, scheduler, and the I/O subsystem ( ``-'' implies feature not supported)}
\label{tab:related_tools}
\resizebox{\linewidth}{!}{
\begin{tabular}{|c|c|c|c|c|}
\hline
\multirow{3}{*}{\bf{Tools/Interaction}} &  \multicolumn{4}{c|}{\bf{MPI Runtime}}\\ \cline{2-5}
    &  \multirow{2}{*}{{\bf Application}} &  {\bf Network} &  \multirow{2}{*}{{\bf Scheduler}} &  {\bf I/O}  \\ 
    &                              &  {\bf Fabric}  &            &  {\bf Subsystem} \\ \hline
{\color[HTML]{009901} \textbf{Proposed \name 
~\cite{gpu-inam,pearc-inam, subramoni2016inam}
}}                            & \color[HTML]{009901}\ding{51}               & \color[HTML]{009901}\ding{51}               & \color[HTML]{009901}\ding{51}               & \color[HTML]{009901}\ding{51}               \\ \hline
TAU~\cite{tau}                                                                     & \ding{51}               & \ding{51}               & -               & \ding{51}               \\ \hline
HPCToolkit~\cite{hpctoolkit}                                                              & \ding{51}               & -               & -               & -               \\ \hline
Intel Vtune~\cite{vtune}                                                             & \ding{51}               & -               & -               & \ding{51}               \\ \hline
IPM~\cite{IPM}                                                                     & \ding{51}               & -               & -               & -               \\ \hline
mpiP~\cite{mpip}                                                                    & \ding{51}               & -               & -               & -               \\ \hline
Intel ITAC~\cite{IntelITAC}                                                             & \ding{51}               & -               & -               & -               \\ \hline
ARM MAP ~\cite{arm-map}                                                                 & \ding{51}               & -               & -               & \ding{51}               \\ \hline
HVProf ~\cite{hvprof}                                                                 & \ding{51}               & -               & -               & -               \\ \hline
PCP~\cite{xdmod} (used by XDMOD)~\cite{PCP}                                                                 & -               & \ding{51}               & \ding{51}               & \ding{51}               \\ \hline
Prometheus~\cite{Prometheus}                                                                & -               & \ding{51}               & \ding{51}               & \ding{51}               \\ \hline
Mellanox FabricIT~\cite{fabricit}                                                                & -               & \ding{51}               & -               & -               \\ \hline
BoxFish ~\cite{boxfish}                                                                & -               & \ding{51}               & -               & -               \\ \hline
LDMS ~\cite{ldms}                                                                   & -               & \ding{51}               & -               & \ding{51}               \\ \hline
Collectl ~\cite{collectc}                                                              & -               & -               & -               & \ding{51}               \\ \hline
Integrated Manager for Lustre ~\cite{iml}                                                                    & -               & -               & -               & \ding{51}               \\ \hline
Lustre Monitoring Tool ~\cite{lmt}                                                                    & -               & -               & -               & \ding{51}               \\ \hline
Darshan ~\cite{darshan}                                                                 & -               & -               & -               & \ding{51}               \\ \hline
\end{tabular}
\vspace{-2.0ex}

}
\end{table}

\begin{figure}[htbp]
    \begin{center}
    \includegraphics[width=1\columnwidth]{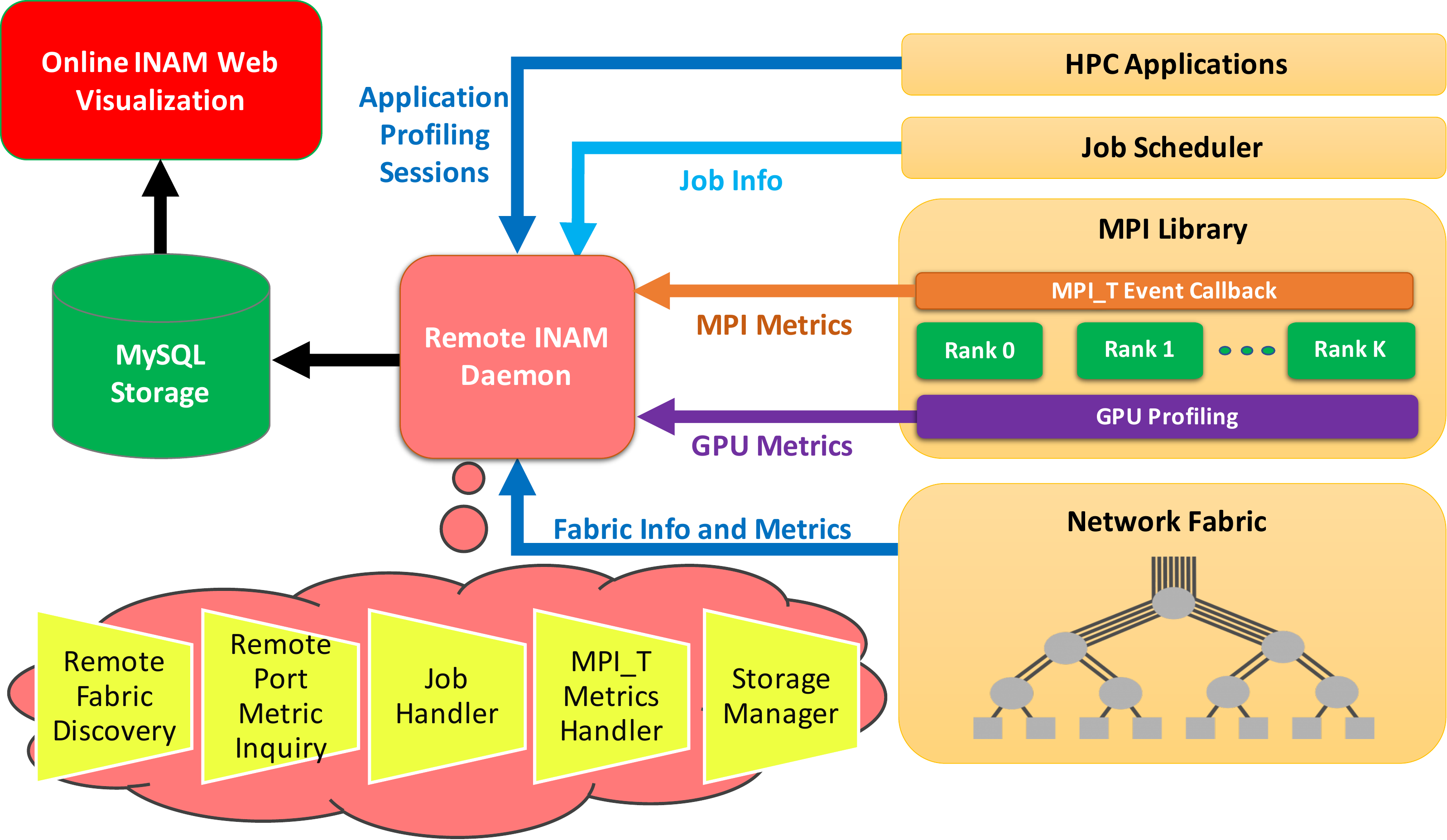}
    \caption{High-level overview of OSU INAM}
    \label{fig:high-view}
    \end{center}
    \vspace{-2ex}
\end{figure}

To fill the gap of a unified, holistic, online, and scalable profiling tool for the HPC communication fabric, we designed and implemented OSU InfiniBand Network Analysis and Monitoring tool - \emph{INAM}. Figure~\ref{fig:high-view} presents an overview of INAM. INAM is capabale of profiling and monitoring large-scale InfiniBand networks with low overhead~\cite{pearc-inam}, profiling GPU and CPU intra-node communication~\cite{gpu-inam}, and profiling both MPI and the job scheduler~\cite{subramoni2016inam}. Therefore, INAM supports profiling HPC/DL applications, MPI communication run-times, and the job scheduler. 

In this paper, we target to propose new visualization designs/features to enable a holistic view of the HPC network communication, and, as a proof of concept, integrate these designs as a new mode into INAM while enabling low-overhead, fine-grained I/O profiling support.

\section{Challenges and Motivation}
\label{motiv}

Profiling tools in the HPC ecosystem are divided into system-level and user-level categories. While system administrators have access to system-level profiling, HPC users face a more limited set of HPC profiling tools. In that context, our design and implementation of INAM targets to bridge this gap and introduce \emph{user-level novelty} in terms of profiling and visualization. There is a need for a more agile, DevOps-oriented approach to HPC. DevOps is a set of practices that combines software development (Dev) and IT operations (Ops). In that sense, we targeted INAM as a DevOps tool for all types of HPC users like domain scientists, HPC software developers, and HPC administrators. INAM provides profiling information for developers and monitoring features for HPC operators.

Usually when HPC users get node allocations, they do not know their job topology. The locality of the allocated nodes of a job can impact the job's performance as there are hundreds if not thousands of jobs running on in-production clusters. Therefore, a node allocation where all the nodes are connected to a switch endures lower performance jitter than a case where the nodes are spread across the cluster and connected by root and edge switches. The communication traffic from the rest of the job/node allocations interplay with each other, specifically on root switches, which determines the performance of the users' jobs. Currently, user options for understanding are limited or require different tools to provide this holistic system-level insight for the breakdown of communication dimensions. Current state-of-art tools fail to provide an understanding of the cluster topology graph with support for link utilization and visualizing a breakdown of communication traffic in an easy-to-understand manner.

The holistic visualization breakdown of communication provides an insightful view into jobs performance on the system-level. For example, HPC job schedulers enlist node allocations for each job, but they do not give characteristics on what each individual job is currently doing and how that impacts the different components of the HPC system. If a job is dumping a lot of data to the file system due to a checkpoint operation or because it has encountered a segmentation fault and is currently in the process of dumping cores, it is going to negatively affect all other processes in the system. Therefore, it is advisable that such ”high-value” jobs be closely monitored by the system administrator. The same is true for the other users to provide novel user-level insight without overwhelming such users with too much information or less relevant details that require a learning curve.

Another aspect to consider is that most of the profiling tools in the HPC community are for \emph{postmortem analysis}. They are used after something goes wrong to figure out the cause. Such analyses can be used to avoid future resource wastage, but the computing resources involved in the faulty job are left underutilized or wasted. Further, if it is a large-scale job, it would impact the system-level performance as explained earlier. The key requirement is to ensure a low overhead and scalable profiling approach for in-production jobs while maintaining fine-grained cross-stack profiling. Furthermore, considering the profiling overhead hides concurrency bugs. Having a low-overhead profiling method, handling, and storing of the data is essential yet challenging. 

\subsection{Problem Statement}
\label{problem}
To better clarify the problem and challenges we ask the following questions and seek to address them by our designs throughout the paper. 
\begin{enumerate}
    \item Can we visualize the details of network topology in an easy-to-understand and interactive manner along with link usage?
    \item How can we present the breakdown of network traffic in terms of communication library, I/O filesystem, InfiniBand unicast and multicast data traffic, along with network topology and link usage to have a holistic and unified visualization?
    \item By using \#1 and \#2, how can we design a tool that enables holistic, low-overhead, scalable, and in-depth understanding of I/O communication traffic on the interconnect through tight integration with the MPI runtime, HPC application, and job scheduler? 
    \item Instead of performing postmortem analysis, can we come up with a HPC DevOps-oriented profiling approach for HPC to increase agility in detecting issues without overwhelming users with less relevant data? 
\end{enumerate}

\subsection{Contributions}
\label{contri}
\begin{figure}[htbp]
   \centering
   \includegraphics[width=1\columnwidth, height=6.5cm, page=1]{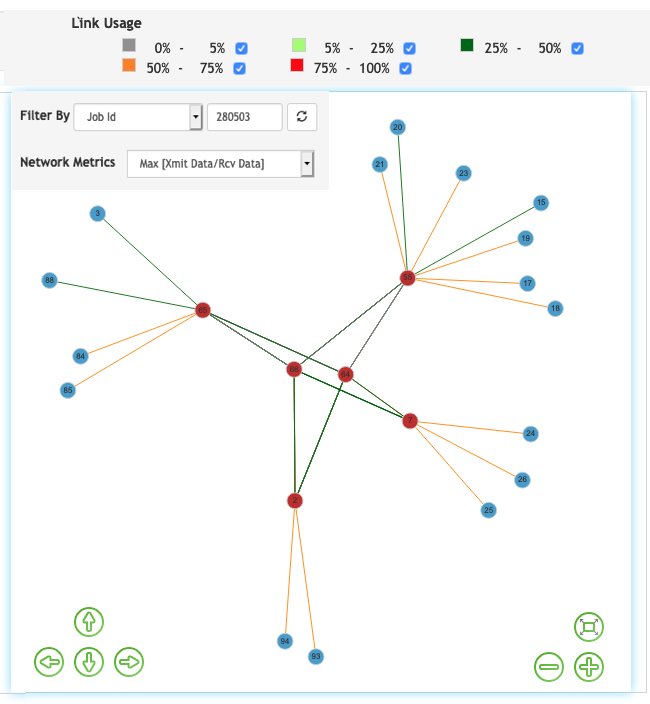}
    \MyCaption{Real-time network view of the job topology for 16 nodes. The blue nodes represent nodes and the red nodes represent switches. Links are colored online based on the network traffic on them. User can select different network metric options like MPI Traffic, I/O traffic to color the links. A historical replay of the job topology is available to visualize the link usage over time for the selected metric.} 
    \label{fig:topo}
\end{figure}

\begin{figure}[htbp]
\centering
\includegraphics[width=\columnwidth]{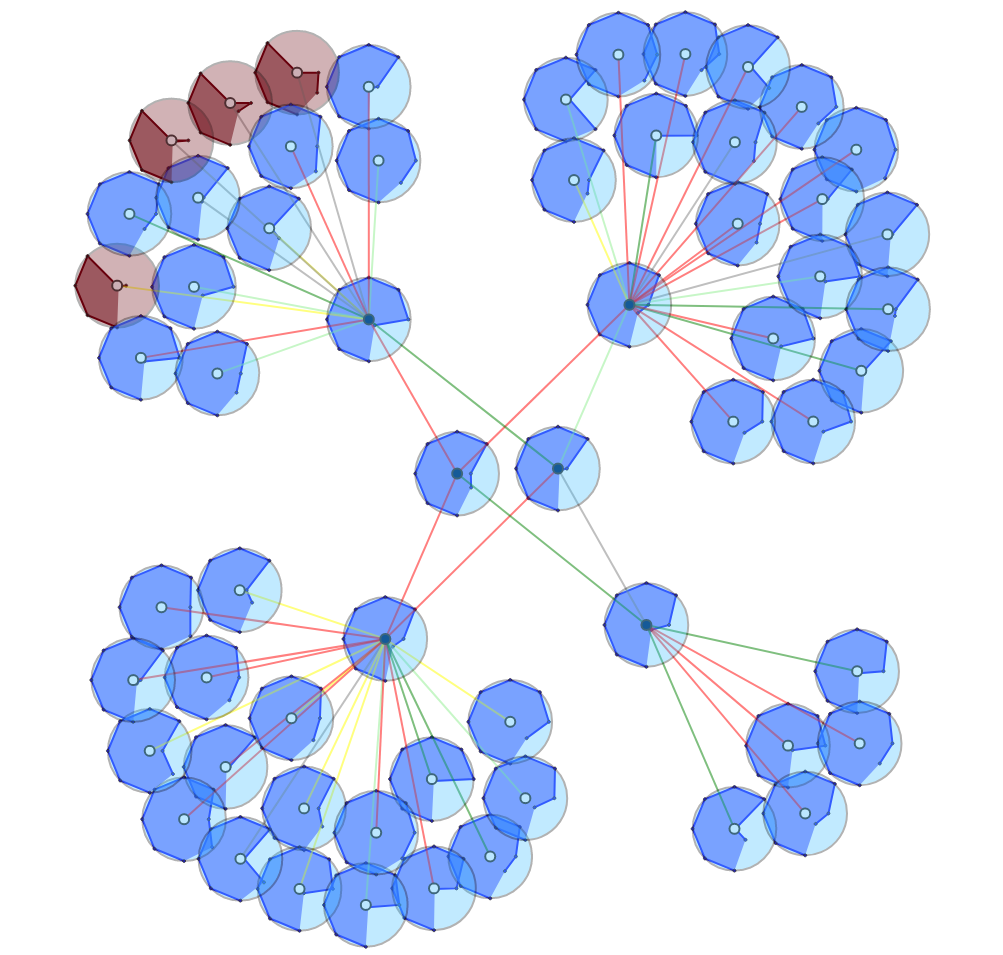}
\MyCaption{Network visualization of communication stack performing MPI AlltoAll with showing RadarPie for all the nodes and switches involved in the job. The nodes that have throttling communication are identified in dark red. The white dots are nodes and the blue dots are switches.}
\label{use-case3}
\end{figure}

\begin{figure}[htbp]
    \centering
    \includegraphics[width=1\linewidth]{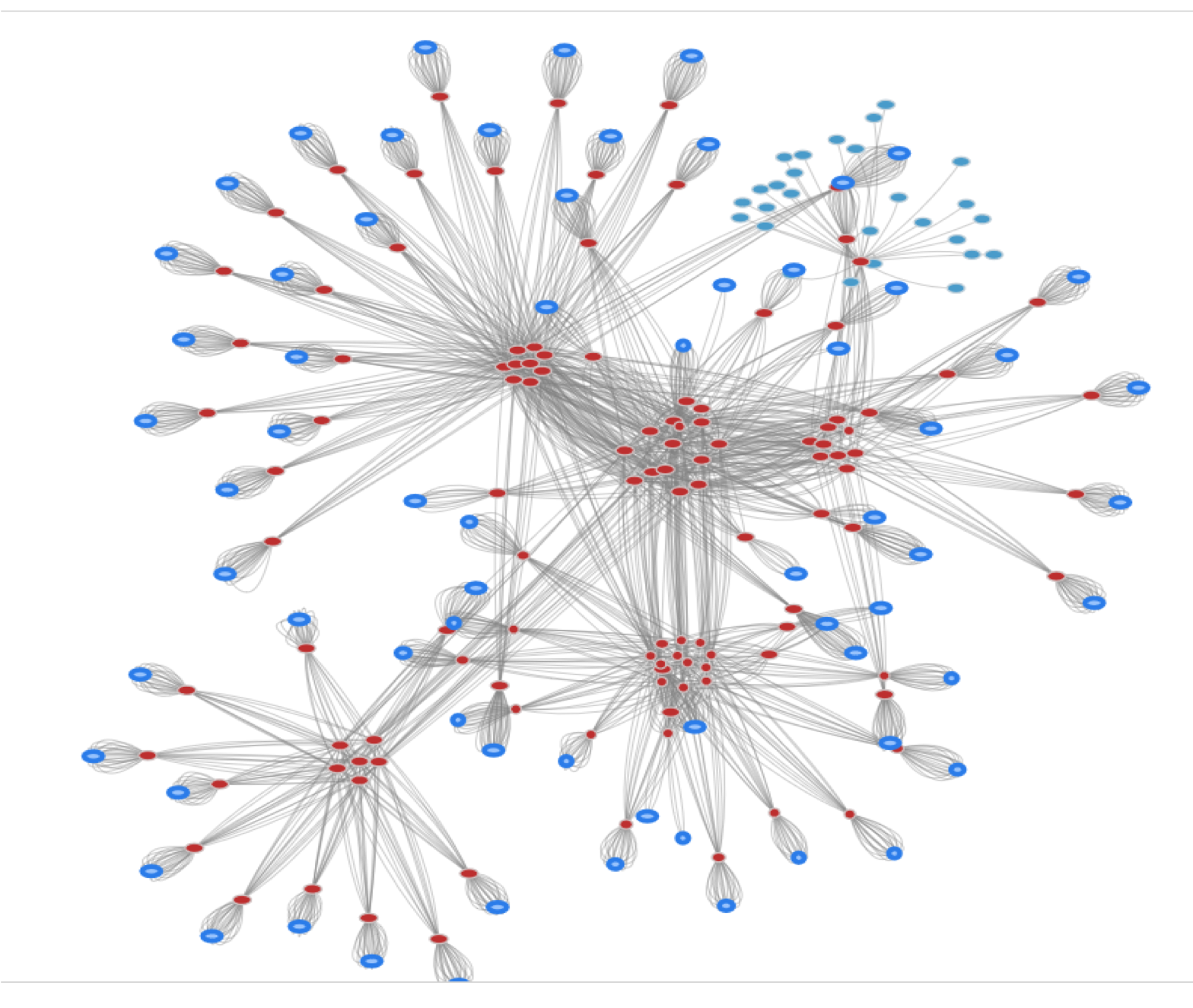}
    \MyCaption{Network topology visualization of Frontera cluster with hidden compute nodes - 8,811 nodes, 494 switches, 22,819 network links - the compute nodes are clustered into one node to be able to show the network topology of Frontera cluster without visualization congestion. The clustered nodes could be expanded as shown in the upper right side of the figure. For full picture please refer to~\cite{inamdownload}.}
    \label{fig:frontera}
\end{figure}

In this paper, we address these challenges and propose three different innovative visualization to enable holistic cross-layer observation and introspection of communication traffic and enable cross-layer network communication profiling in real-time with high-fidelity through \name. \textbf{To summarize, the key contributions of
this paper are as follows:}
\begin{itemize}
    \item We propose and design three holistic communication visualization capable of showing different profiling sources from network fabric counters, communication library, and I/O file system for HPC systems.
    \item We evaluate our proposed visualization performance and show scalability up to 8,811 nodes and 494 switches on Frontera supercomputer (Figure \ref{fig:frontera}).
    \item{Design and implement the proposed extensions in a
    scalable fashion with low overhead as a new feature in \name for a proof-of-concept.} 
    \item{Redesign the MPI library to collect I/O statistics online with low overhead.}
    \item We propose \name, a cross-stack profiling tool to aggregate, correlate, and visualize profile data from different profiling sources independently from other profiling infrastructures and construct a holistic view of HPC communication stack.
\end{itemize}
As part of new profiling visualization, we show detailed and holistic communication statistics in Figures \ref{fig:topo}, \ref{Ri2-multi}, and \ref{use-case3}.

The rest of the paper is organized as follows. Section \ref{sec:back} give background on HPC interconnect, INAM, and Lustre file system. Section \ref{viz} presents our proposed visualization design to enable holistic view of communication stack. Section \ref{sec:design_lustre} describes our implementation designs to gather I/O and correlate it with MPI and network topology. Section \ref{sec:usecase} showcases the benefits and motivating examples of some insight that are enabled by our implementation. Section \ref{sec:perf} further evaluates the overhead of the proposed designs. Section \ref{sec:related_work} discuss the related work in the community. Section \ref{sec:conclusion} concludes the paper.
\MySection{Background}
\label{sec:back}

\MySubsection{HPC Interconnect Technology}
\label{sec:IB}

High Speed Ethernet, InfiniBand and Omni-Path are the major interconnects used in the modern HPC community. InfiniBand is a high speed and high bandwidth switch interconnect providing low latency, and is used in more than 32\% of the top500~\cite{TOP500} list including \#2 \textit{Summit}, \#3 \textit{Sierra}, and \#5 \textit{Selene}. The InfiniBand interconnect provides more than 40\% of the performance share of the top500 list. In fact, 7 out of the top10 supercomputers are using an InfiniBand-family interconnect. The latest InfiniBand HDR adapters offer a bidirectional bandwidth of 200Gbps. The data collected for each link are gathered from libibverbs provided by Mellanox~\cite{mellonox-switch}. The data shows the total number of packets sent/received, the total number of bytes sent/received, and a classification of the number of unicast and multicast traffic in bytes for sent/received. Typical users cannot access this information since it requires root access. In this section we provide the necessary background for the paper. We define two types of switches: 
\begin{itemize}
    \item Edge switch: where the switch has links connected to compute nodes. 
    \item Root switch: where the switch is not connected to the compute nodes and plays an infrastructure role in connecting the other edge switches together. 
\end{itemize}

\MySubsection{OSU INAM}
\label{sec:INAM}
OSU INAM is a network profiling, monitoring, and analysis tool designed to bridge the gap mentioned in Section \ref{sec:intro} by providing real-time and scalable insights into the  communication traffic of HPC interconnects. Further, INAM supports GPUs and is tightly integrated with MPI runtime and the job scheduler. It provides insights and profiles for various HPC users like administrators, software developers, and domain scientists. INAM users have the capability to analyze and profile node-level, job-level, and process-level activities for MPI communication. INAM is capable of gathering and storing performance counters at sub-second granularity for very large clusters ($\approx 2,000~nodes$)~\cite{pearc-inam}. It supports gathering metrics from the PBS and SLURM job schedulers. INAM has been deployed at various HPC clusters and downloaded more than 4,200 times from the project website.


\MySubsection{Lustre I/O}
\label{bgnd:lustre}
Lustre is a type of parallel distributed file system that is generally used for large-scale HPC.
Lustre exposes statistics on data transfers for each component
through the kernel-exposed file system. The Lustre file system has various components, such has Metadata
Servers (MDS), in which one MDS per file
system manages one metadata target (MDT). Each MDT stores file
metadata such as file names, directory structures, and access
permissions. The MDS controls the allocation of storage objects
on the Object Storage Servers (OSS). One or more OSS store
file data on one or more object storage targets (OST) - block
storage devices. OSS exposes block devices and serves data to the
Object Storage Client (OSC). The Object Storage Target (OST) is 
responsible for storing the data objects. 
The Object Storage Servers (OSS) manage the OSTs, and there can be a one-to-many mapping.
From the OSS, reading and
extracting data for each client is straightforward since the ``stats''
files report data in finer granularity and include client
details. The server readily knows the traffic route since it is
aware of the client details and the data transfer statistics. However, 
it requires a daemon to be running on the server at all times, 
which could cause interruption or jitter in the Lustre service. 
The I/O performance of Lustre has a critical impact on applications and has attracted broad attention. 

\MySection{Holistic Network Traffic Visualizations}
\label{viz}
In this section, we tackle Problems \#1 and \#2 as discussed in Section~\ref{problem}.
We use a graph view to show the network topology, which provides users an understanding of their job topology. We provided two versions to show the network topology. In particular, we started with a simplified version of the network by having a single link between pairs of nodes and switches and moved to a multi-link network visualization to show the entirety of the network in case the user is interested in viewing links between root switches. Finally, we discuss our third option to overlay on top of the first two designs. The visualization designs mentioned in this section are portable to any network communication tool.

\MySubsection{Design \#1 - Single Link Network Topology}
This design uses a single edge between switches to represent the link connectivity. Therefore, it indicates logical connectivity of switches to each other. Since there is one physical link between edge switches and nodes, the (node, switch) links will be the same.
On top of providing the backbone structure of the network, the colors of the links serve as an indicator to show the utilization rate. The utilization for the links between switches is aggregated and colored. This gives users flexibility in choosing their desired visualization and does not overwhelm them with extra link data. We used a color scale to indicate that the best is at 50\% link usage and moving data near the link's capacity is not desired since it results in communication bottlenecks. 
Figure~\ref{fig:topo} depicts the single link visualization design displaying and filtering a specific 16-node job. The visualization shows the node Local Identifiers (LID). Hovering over the nodes shows the job ID, node name, and node Global Identifier (GUID). The users can hover the mouse on different elements of the graph to see messages about the IDs of links and nodes, the source and destination of specific links, the job ID of the running traffic on the nodes, and the detailed utilization number of each link. This design of detailed artifacts supporting the overall topology view provides a better user experience with a high level of interaction.

\MySubsection{Design \#2 - Multi Link Network Topology}

\begin{figure}[htbp]
   \centering
    \includegraphics[width=1.0\columnwidth, ]{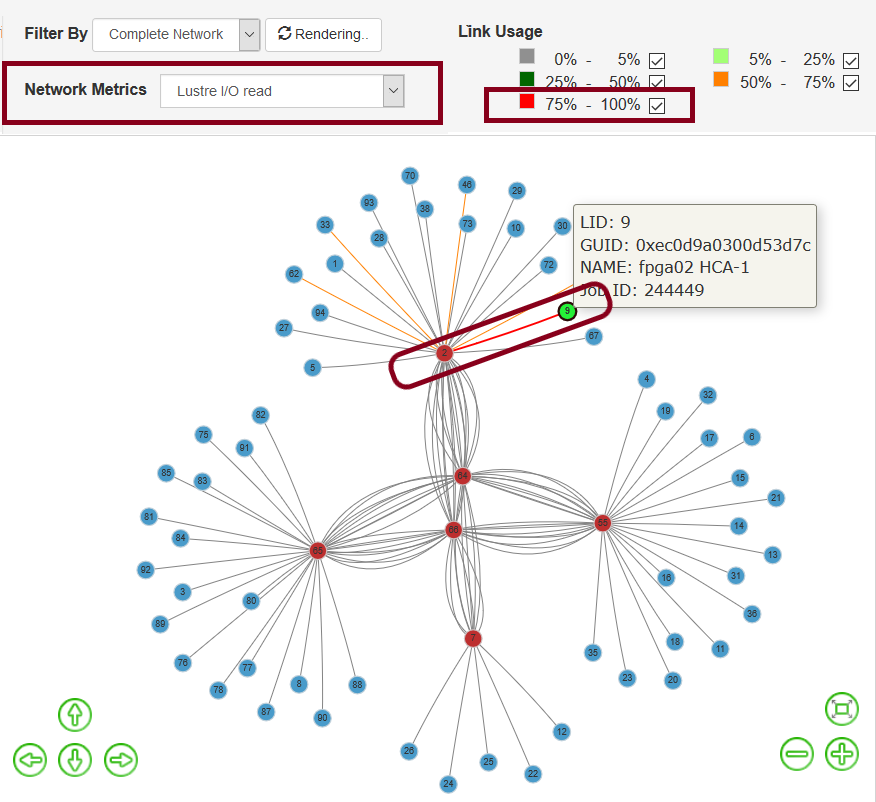}
    \MyCaption{Network view - Link utilization by Lustre traffic. Lustre traffic on the highlighted link is between 50\%-70\% of the link capacity}
    \label{fig:lustre-traffic}
\end{figure}

We propose a multi-link visualization that preserves the coloring strategy and hovering details like the single link network topology, depicted in Figure~\ref{fig:lustre-traffic}, to allow monitoring of the link usage where the use of multiple links between two switches matters. This visualization is better suited for system-level analysis. This is useful for understanding the link usage between root switches where multiple paths exist. We also decided to curve the links to help differentiate different links. We implemented our curve interpolation algorithm as an option for this visualization as shown in Figure \ref{Ri2-multi}.
Let two ends of the edge be $A(x_1, y_1), B(x_2, y_2)$, suppose we have $n$ links between two nodes, for all $k = 1,..,n$, we draw the spline that passes through $[A, (x_1+\frac{k(x_2-x_1}{n}, y_2+\frac{k(y_1-y_2)}{n}), B]$. The users can easily identify the root nodes connecting to the switches while still comprehending the link usage statistics on each distinguishing link.

While Figures~\ref{fig:topo} and \ref{fig:lustre-traffic} depict the job topology with link usage, they are limited to show one network metric at a time, which makes it harder to compare different communication metrics such as I/O and MPI traffic on the network fabric. In such cases, users need to change metrics or render a new visualization to compare each metric of communication. To address this issue we suggest using a parallel coordinate chart for communication traffic.

\begin{figure}[ht]
\vspace{-2ex}
\centering
\includegraphics[width=\linewidth]{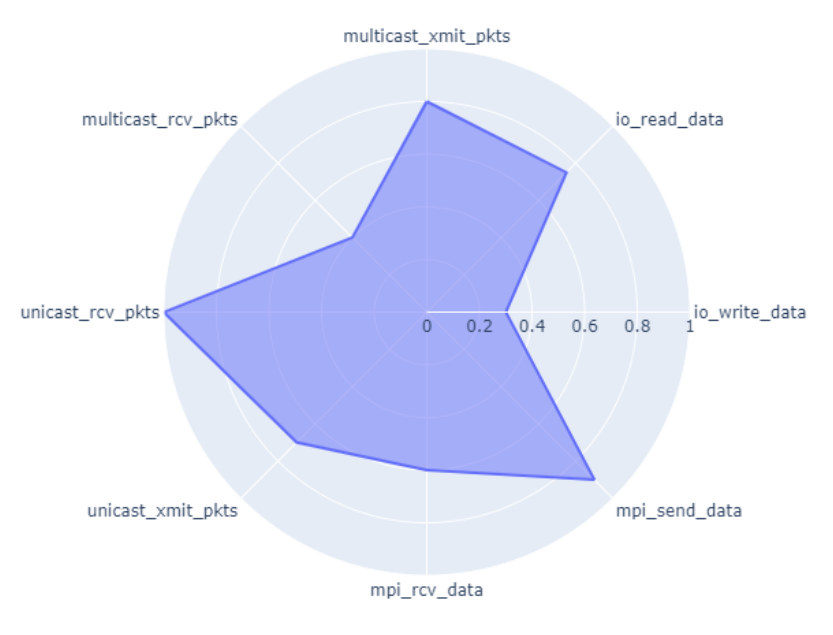}
\MyCaption{Explanation and mapping of network metrics to parallel axes in a circular view used for overlaying switches.}
\label{fig:RadarPie}
\vspace{-2ex}
\end{figure}

\begin{figure}[htbp]
\centering
\includegraphics[width=\linewidth]{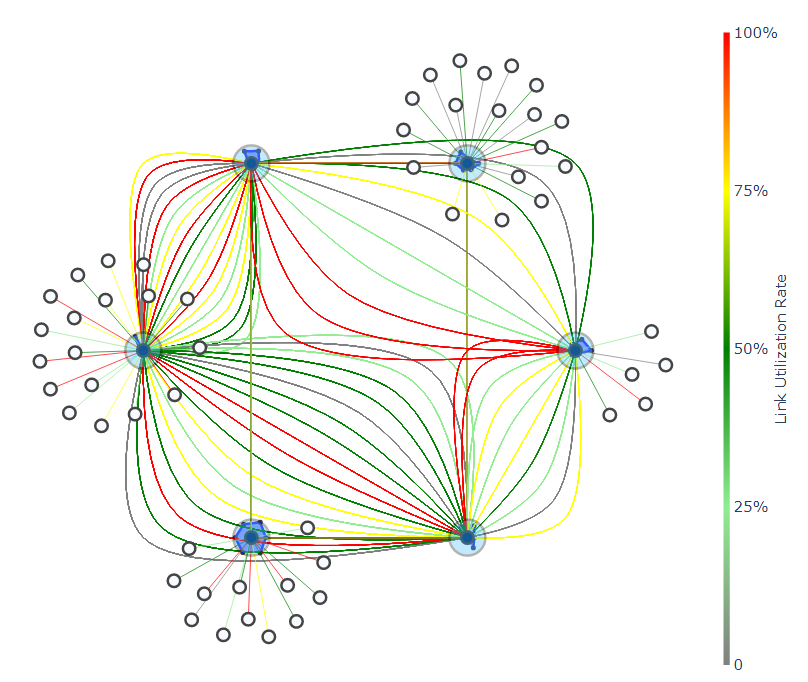}
\MyCaption{Multi Link RadarPie topology view of RI cluster with colored links based on their usage on the right side of the figure and customized curving of switch to switch links to increase readablity. Green indicated good usage and red indicates link congestion. The RadarPie axes are shown in Figure \ref{fig:RadarPie}.}
\label{Ri2-multi}
\end{figure}

\MySubsection{Design \#3 - RadarPie Visualization}
As Problem \#3 discussed in Section~\ref{problem}, visualizing the breakdown of network metrics and comparing them in one visualization is a challenging task due to the many dimensions of communication traffic. Parallel coordinate chart~\cite{parallel-cor} is a common way of visualizing and analyzing high-dimensional datasets. To show a set of points in an n-dimensional space, a backdrop is drawn consisting of n parallel lines, typically vertical and equally spaced.
We use the idea of parallel coordinates to overlay on the top of nodes in Figure~\ref{fig:topo} in a circular fashion depicted in Figure \ref{fig:RadarPie}, called \textbf{RadarPie}.

\begin{figure}[htbp]
\centering
\includegraphics[width=\linewidth]{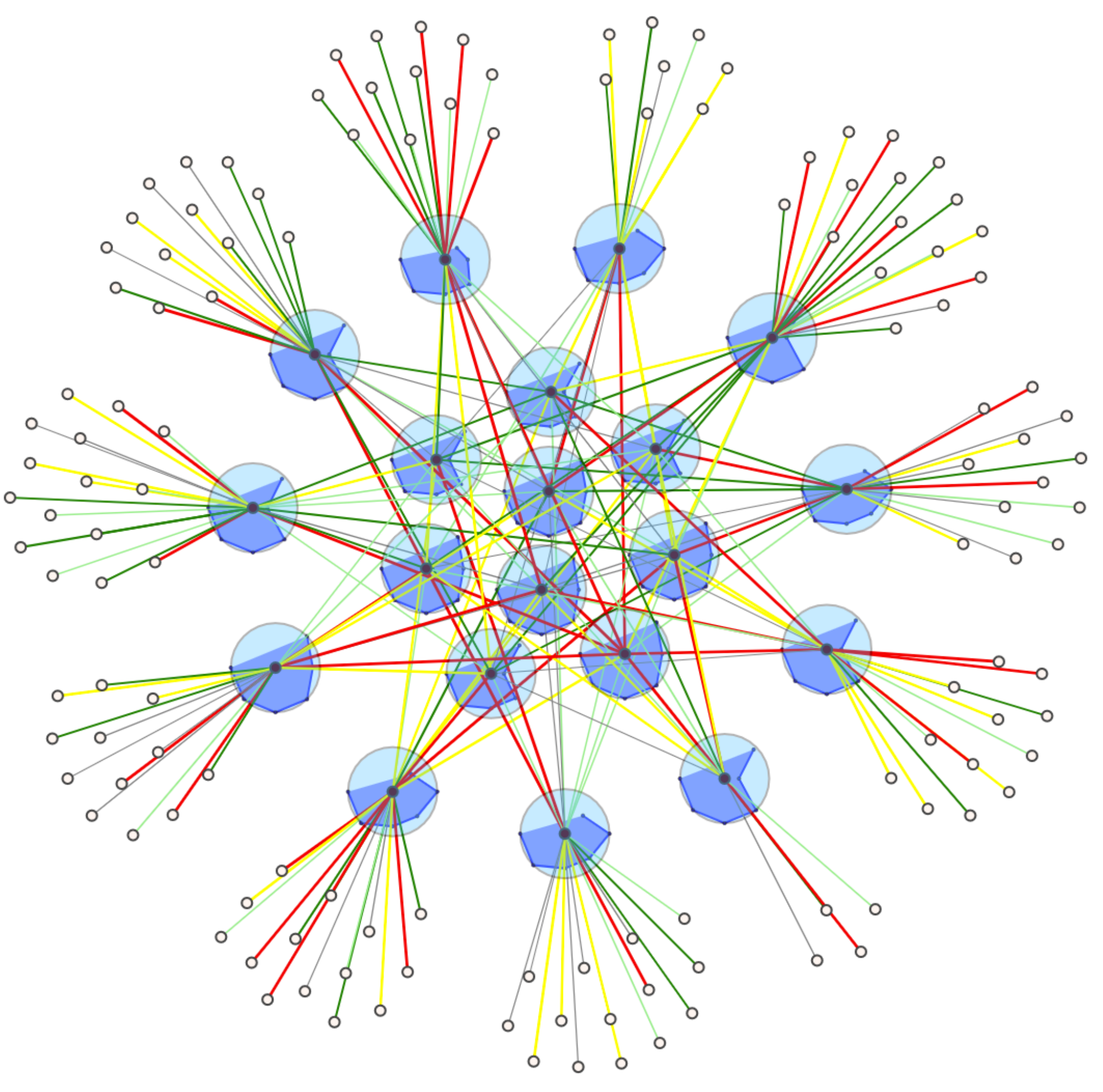}
\MyCaption{Single link RadarPie topology view of RI cluster with colored links based on their usage. Green indicated good usage and red indicates link congestion. The RadarPie axes are shown in Figure \ref{fig:RadarPie}. The root switches are in the center. }
\label{Ri2-single}
\end{figure}

By this design, we visualize all the dimensions of network traffic like InfiniBand unicast and multicast data, MPI-level data, and I/O-level data along with network topology and link usage to present a holistic visualization. For the network nodes, we display a full breakdown of network traffic in terms of (multicast,unicast, MPI, I/O) data bytes sent/recv. This gives us 8 different axes to insert into circular parallel coordinates. For ease of understanding and space, only the RadarPie charts for switches are shown in Figures \ref{Ri2-single} and \ref{Ri2-multi}. The user can select the number of axes to show for the RadarPie.

The RadarPie chart supports two modes of absolute and relative values. For the absolute mode, the value attributes are normalized with the link bandwidth. The relative mode uses total sent/received data from reading InfiniBand port counters to perform normalization of other attributes. This is useful to understand the restiveness of communication for a specific job and to figure out if a job is I/O-bound or MPI communication-bound. Another example would compare the multicast vs unicast data.

By using D3 visualization, we ensure interactivity, zooming, and hovering over data points to view further details like absolute values. To include this additional traffic information all within the same visualization, we opted to overlay radar charts on top of each switch node on the graph. By including this traffic information this way, viewers can easily understand the traffic going through one of the switches just at a glance and compare switches to each other by comparing the shape of the radars. To get the exact number or percentage of a specific axis attribute, the user can hover over part of the radar chart to get the exact figure. This capability enables users to compare the different metrics of a network and compare them to total bytes sent/received on the link without having to create a time-series graph. Moreover, RadarPie chart enables the easy visual comparison of node traffic.


\MySection{Introspecting MPI and I/O Traffic Together}
\label{sec:design_lustre}

In this section, we discuss the designs for introspecting and evaluating the MPI runtime and the I/O together with low overhead to tackle problem \#3 in Section~\ref{problem}. Note that, although this initial design was done for Lustre, similar schemes apply to other file systems like BeeGFS. 

\MySubsection{Gathering I/O Traffic}
\label{sec:design_Lustre_gather}

\begin{figure}[htbp]
   \centering
   \includegraphics[width=1\linewidth]{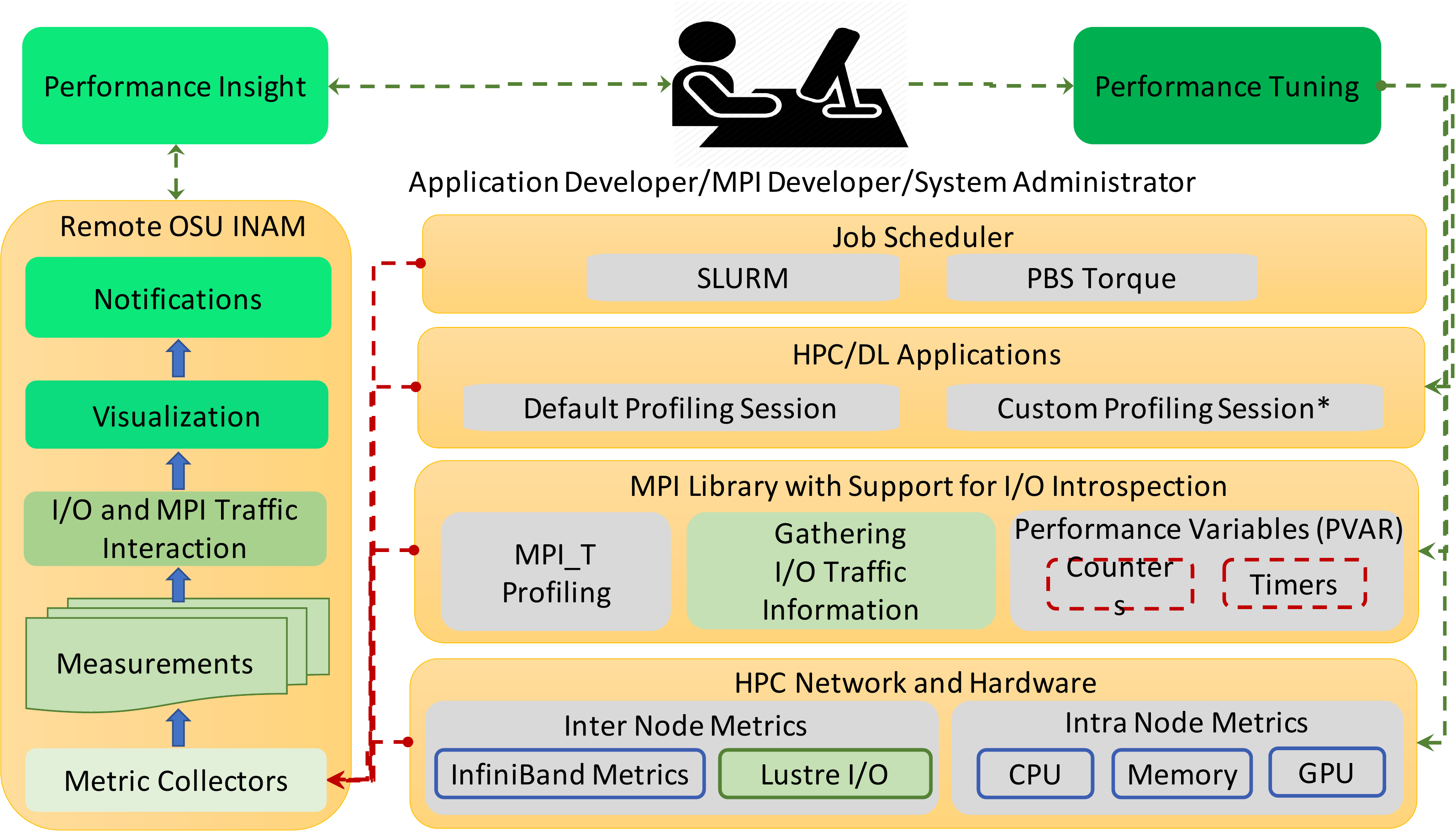}
   \MyCaption{Detailed design and the flow of \name - Red arrows shows the flow of gathering metrics from HPC stack on right, blue arrows show the high-level internal flow of \name for processing metrics across profiling stack, and the green arrows flow of applying insight and performance recommendation to their application. The green boxes are new or modified.}
    \label{fig:detailed_design}
    \vspace{-2ex}
\end{figure}

OSU INAM monitors MPI information using MPI\_T as discussed in~\cite{subramoni2016inam}. 
The MPI profiling metrics include the number of bytes sent and received by each algorithm for each MPI rank, the
number of MPI calls invoked, and time taken by each MPI primitive
and its algorithms for each MPI rank.
The MPI library was modified to collect Lustre I/O data at the
client side at run-time. The following four data points on reads and
writes are collected for each Object Storage Target (OST) in each client: ``number of times the read/write event
happened'', the ``minimum bytes read/written in one event'', the
``maximum bytes read/written in one event'' and the ``sum of bytes
read/written''. Additionally, we collect the OST name and its
corresponding Object Storage Servers (OSS)'s IP address.
The MPI library then sends this information to
\name every user-defined Lustre time interval through the
InfiniBand UD channel by using the framework discussed in~\cite{subramoni2016inam}.
The I/O profiling frequency is by default the same as the MPI profiling frequency, which
allows the synchronization of I/O and MPI traffic timestamp records. The remote INAM daemon collects and processes this information in a parallel and efficient manner described in \cite{pearc-inam}, and stores the measurements into a MySQL database. 

As mentioned above, we need to ensure the low-overhead profiling of I/O.
Lustre Traffic is linear with respect to the number of OST. Considering each Lustre storage record is 311Bytes, for Lustre traffic we have: 
\begin{gather}
\label{eq2}
\text{Traffic}_{Lustre} = \text{NumProcs} * \text{RecSize}(311B)  \\ \notag
* \text{NumOSTs} * \text{Frequency} 
\end{gather}
Given that HPC clusters are capable of 56Gbps to 200Gbps this is an insignificant amount of data and ensures a low-overhead profiling approach. 

The user also has the
flexibility to look at various servers present in a
node/job. This information is available
at both the node-level and job-level granularities. With this setup, the user can monitor each job’s specific communication pattern for the Lustre server
and the route through the network.

\MySubsection{Correlation \& Evaluation of MPI and I/O Traffic}
\label{sec:design_Lustre_introspect} 
\name's data collection interface intercepts the data received from MPI processes
and stores measurements in a persistent data store as shown in Figure~\ref{fig:detailed_design} on a remote \name node. 
While MPI processes provide information on the IP address to which the
Lustre traffic was directed, we need the corresponding node name
to correlate Lustre traffic with MPI traffic in the cluster.
To achieve this, we use the following two techniques to identify the IP-node name
mapping: a) Address Resolution Protocol (ARP)-based Scheme, and b) Reading the IP to node mapping from the \texttt{/etc/hosts} on the compute node.
We primarily use \texttt{/etc/hosts} method to get the mapping of the IP
addresses to the hostnames and rely on the ARP-based mechanism as a fallback approach.

To support introspecting the MPI library along with the I/O subsystem, Section~\ref{sec:intro}, \name remotely gathers fabric discovery and routing information periodically~\cite{pearc-inam}. Since HPC systems have static routing, the change in the routing and fabric is minimal. Once the details of Lustre servers-nodes are known, \name uses fabric information to find the path of the I/O traffic and MPI traffic using route table. Considering we are gathering this information in real-time, we know how much traffic for MPI and I/O is going through each link and can see if they are colliding in the network.

Using the Lustre I/O information collected by each MPI process,
we can identify a) the Lustre traffic generated by each node, b)
the list of nodes to which data was read/written, and c) the
information on the job that is inducing Lustre traffic. \name can
then leverage the route and Lustre information as mentioned above
to identify the path taken by each job's Lustre traffic. 
With this information, \name can
correlate Lustre traffic with MPI traffic to identify points of
contention in the network fabric system where I/O traffic affects
the performance of MPI communication.

With the information collected from Lustre I/O traffic, we
render the percentage of link bandwidth utilized by I/O traffic
for each link in the network. 
We also render the percentage of link bandwidth utilized by MPI traffic. This information 
together enables system administrators to compare and
identifies the points within the system dominated by I/O traffic.

\begin{figure*}[t]
   \centering
    \includegraphics[width=0.9\linewidth]{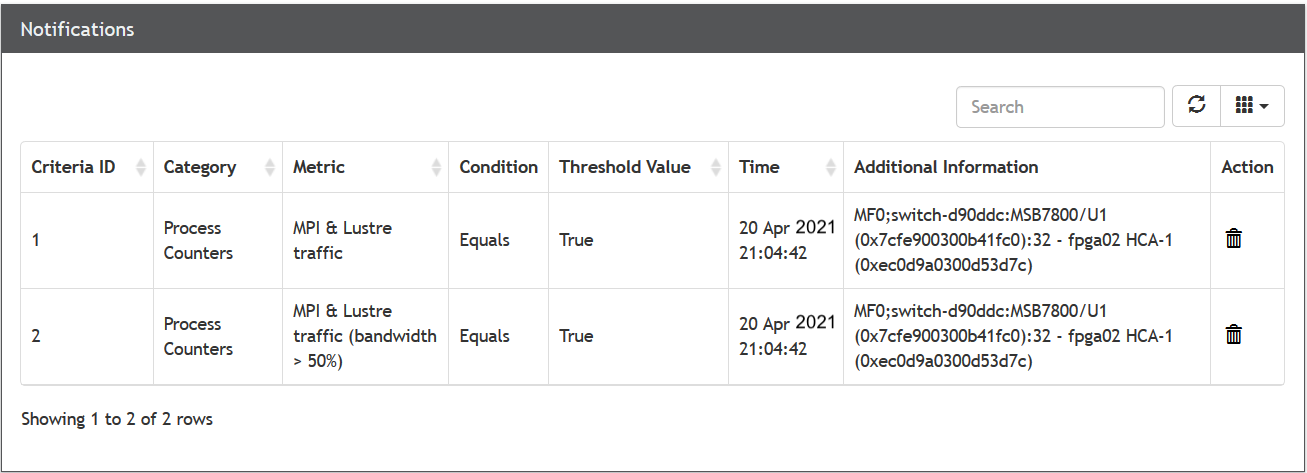}
    \MyCaption{In-production web interface to track events and notifications generated by \name that occurred in the past. We see that MPI Traffic and Lustre Traffic shared links specified in additional information}
    \label{fig:notif_ui}
\end{figure*}

\MySubsection{Performance Notifications}
As a part of tackling problem \#4 in Section~\ref{problem},
we have a notification system in place which looks
for user-configurable events at user-specified time intervals and
notifies the users(like system administrator) via email. This helps the system administrator to monitor and operate the HPC system. As shown in Figure \ref{fig:notif_ui}, we also have a
web-based user interface that helps users
keep track of events. \name has
support to report notifications when metrics such as
``LinkDowned'', ``XmtDiscards'', ``RcvErrors'', ``VL15Dropped'', ``Bytes
sent'', ``Bytes received'', ``Link utilization'' exceed/drop
below/equals a user-specified threshold value. Additionally, we
have support to report when MPI \& Lustre traffic co-exist in the
same link which can impact the performance of the job based on a user-specified threshold value.
(Figures~\ref{fig:lustre-traffic}). With this quiet, simple, and useful support of performance notifications, INAM users can perform online analyses, which shortens debugging time and saves resources spent detecting system problems.

\MySection{Usage Scenarios}
\label{sec:usecase}

\MySubsection{System level Impact of I/O and MPI}
This use case is to see the insight of using the new visualization on the system-level to detect exist of MPI and I/O traffic in the root switches.
For this case, we run 144 nodes to perform MPI AlltoAll on RI cluster. MPI AlltoAll is a communication-intensive operation where all processes send the same amount of data to each other and receive the same amount of data from each other. We observed lower than expected performance for the jobs on RI cluster with up to 20\% performance variation among different iterations of running the job. Upon visualizing RI full-stack network traffic, as shown in Figure \ref{Ri2-single}, we noticed that there is a significant amount of I/O traffic on root switches even though we are running on 94\% of the available nodes in the cluster. All root switches are showing I/O read operation. Upon checking the jobs, we noticed that there is another I/O intensive checkpoint restart job injecting the I/O traffic into the system causing the performance variation.

\begin{figure}[htbp]
\centering
\includegraphics[width=\linewidth]{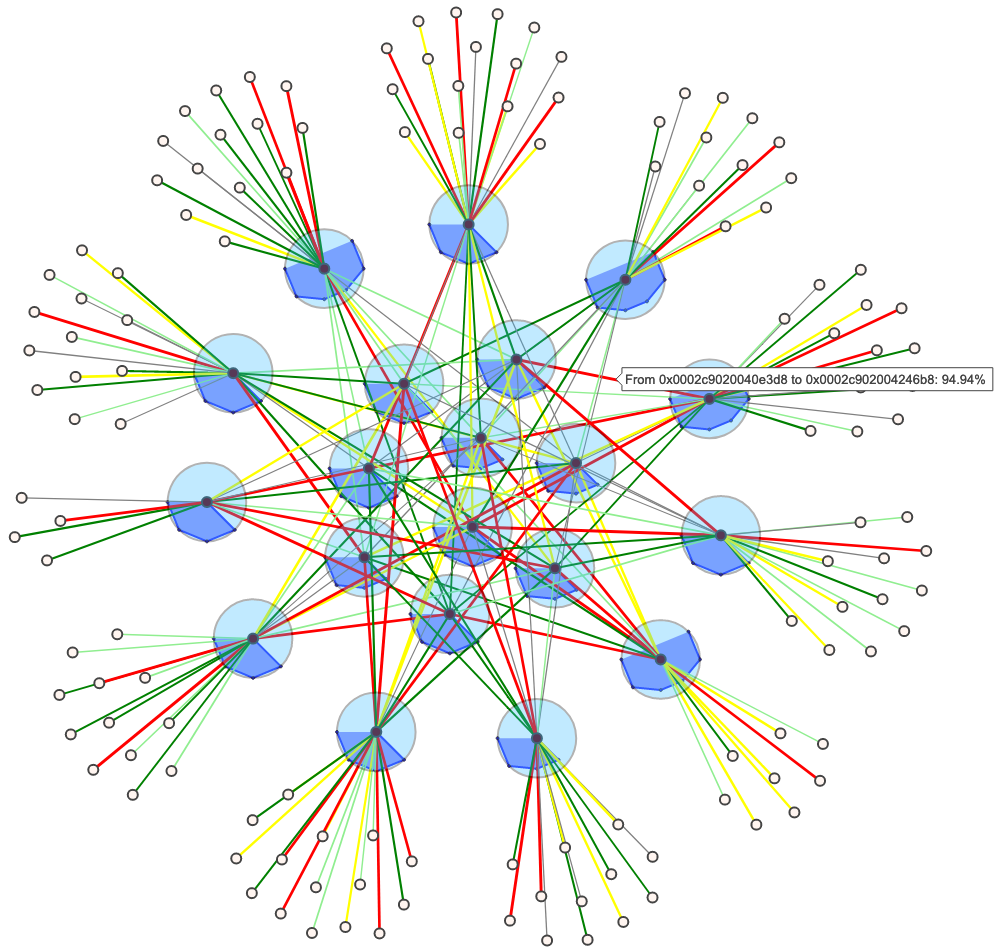}
\MyCaption{System level view of running MPI AlltoAll on 144 nodes. Note that compared to Figure \ref{Ri2-single} there is not I/O traffic on the root switches.}
\label{use-case}
\end{figure}

After waiting for the I/O job to finish, we ran our experiments again and visualized the network traffic while the jobs were running as shown in Figure \ref{use-case}. This time we noticed although there are a few edge switches that have I/O traffic, but the root switches in the center are not impacted by I/O traffic in the system. Comparing Figures \ref{Ri2-multi} and \ref{use-case} shows the impact of locality of I/O traffic on communication performance in the network and how users could quickly detect the existence of I/O in the root switches. 

\begin{figure}[htbp]
\vspace{-2ex}
\centering
\includegraphics[width=\linewidth, height = 6cm]{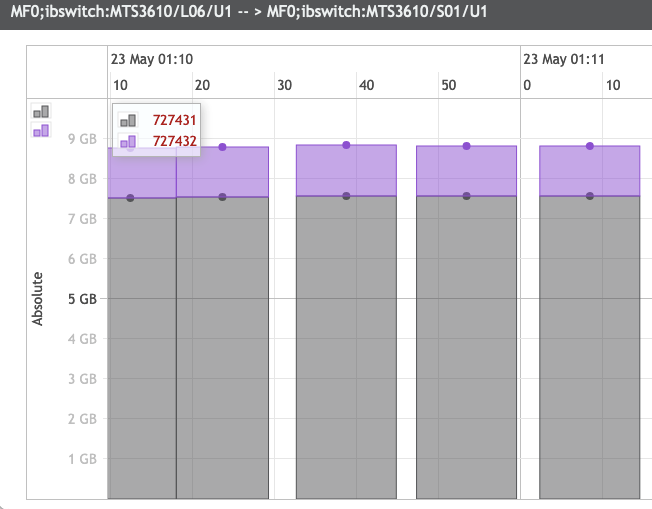}
\MyCaption{A summary of a single link sharing traffic between two jobs resulting in performance interplay between the running jobs.}
\label{link-comp}
\end{figure}

\MySubsection{Two Jobs Sharing a Link}
Another use case scenario may arise when multiple jobs map to the same link between root switches. INAM exploits the paths between every set of nodes in the network since in HPC systems the routing between nodes is static. By using the notification system, we set events to get notification when there is MPI and I/O traffic beyond 50\% on network links used by our job allocation. We expect some jobs to affect each other's performance by congesting the root link, and target to show the breakdown of jobs communication behavior with the designs described above in a systematic manner. First, to detect the links that the usage is within 75\% to 100\% (red color) by using Figure \ref{Ri2-multi}. Then, visualizing the amount of traffic each job transferring on those links in real-time. We notice that Job ID 727431 and Job 727432 are running on the same link. Application users and HPC systems administrators can monitor job interference in real-time and detect the source of link congestion in terms of each job. A Similar view like Figure \ref{link-comp} is available for viewing share of communication traffic on a link for MPI processes of a MPI job.\cite{inamdownload}. 

\MySubsection{Detecting Nodes with Communication Performance Degradation}
The third usage case includes detecting abnormal behavior of communication in a specific job. Figure \ref{use-case3} shows the communication visualization of nodes and switches of a job doing MPI AlltoAll collective operation. We notice that for the nodes shown in dark red the communication pattern of the RadarPie is different from other compute nodes. The user can compare the full-stack communication characteristics across nodes and identify the abnormal communication behavior using this visualization quickly. By detecting the nodes showing lower communication, we can identify the source and location of communication degradation. For example, here we can what layer in the job and what nodes are the bottleneck and resolve it. In this case, the user can report the throttling issue to the system administrator to be evaluated. 

\MySection{Performance Evaluation}
\label{sec:perf}
In this section, we evaluate the visualization designs and the overhead of I/O profiling.

\MySubsection{Experimental Setup}
We conducted our experiments on three clusters at The Ohio State University (OSU) and the Ohio Supercomputer Center~\cite{osc}. Table~\ref{Tab:cluster-spec} summarizes the specification of each the clusters. 

\begin{table}[hbpt]
\caption{The hardware specification of clusters used for experiments}
\label{Tab:cluster-spec}
\begin{tabular}{|l|c|c|c|}
\hline
& \bf RI2         & \bf RI           &\bf OSC \\ \hline
\#Nodes        & 56      & 152         & 1,738\\ \hline
\#Links        & 189     & 542         & 3,579  \\ \hline
\#Switches     & 6      & 20        & 109 \\ \hline
CPU   &  Xeon 2.40GHz   &Xeon 2.53Ghz   &  Xeon 2.50GHz  \\ \hline
\begin{tabular}[c]{@{}l@{}}Cores/node\\ for INAM\end{tabular}& 28       & 8           &24   \\ \hline
Memory/node   & 128GB       & 12GB        & 128GB  \\ \hline

\begin{tabular}[c]{@{}l@{}}Switch\\ Technology\end{tabular}& Mellanox EDR     & Mellanox QDR    & Mellanox EDR   \\ \hline
\end{tabular}
\end{table}

\MySubsection{Visualization Performance Analysis}

\begin{figure}[htbp]
    \centering
    \includegraphics[width=1\linewidth]{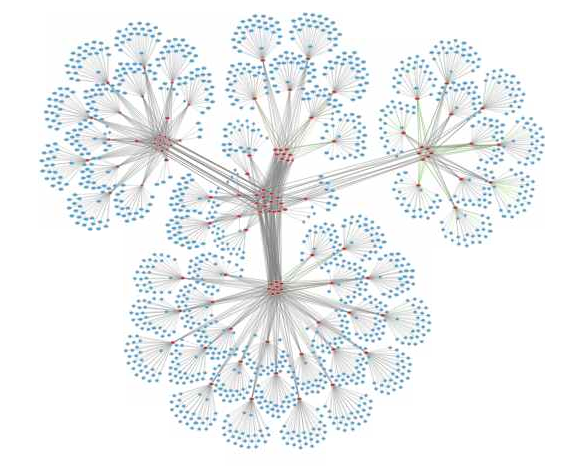}
    \MyCaption{Network topology visualization of OSC cluster - 1,738 nodes, 109 switches, 3,579 network links. For full picture please refer to \cite{inamdownload}.}
    \label{fig:osc}
\end{figure}

In Table \ref{Tab:owens}, we present the latency
of visualization designs for multi-link network topology (Design \#2) and RadarPie on top of multi-link network topology (Design \#3 + Design \#2) for each system with varying number of nodes to visualize. In particular, we move to a larger cluster of nodes(OSC) and links, with 30x nodes and 20x links compared to RI2 cluster to study the impact of number of nodes and links on the latency of visualizations. This shows us that Visualization design \#2 (Figures \ref{fig:lustre-traffic}) provides faster rendering. This is due to caching of network layout. The RadarPie visualization is more computationally heavy and includes retrieving data from various tables in database and overlaying them. RadarPie visualization (Design \#3) takes less than 30 seconds for a cluster size of ($\approx 2,000~nodes$) to show new visualization for all the metrics across all nodes. Figures \ref{fig:osc} and \ref{fig:frontera} show the visualization design \#2 for OSC and Frontera supercomputers. Design \#2 has been publicly released and available to use in INAM\cite{inamdownload}. 

\begin{table}[htbp]
\centering
\vspace*{-1.0\baselineskip}
\caption{Performance evaluation of proposed network visualizations across various HPC clusters}
\label{Tab:owens}
\begin{tabular}{|c|c|c|c|c|c|}
\hline
\textbf{}             & \textbf{RI2} & \textbf{RI} & \textbf{OSC}   \\ \hline
\textbf{Visualization Design \#2}    & 40 ms    & 53 ms      & 206 ms      \\ \hline
\textbf{Visualization Design \#3} & 5.03 s     & 4.85 s      & 26.12 s         \\ \hline
\end{tabular}%
\vspace*{-1.0\baselineskip}
\end{table}

\MySubsection{I/O Profiling Analysis}

In this section, the overheads of profiling for I/O component is evaluated. To be fair and aligned with problem \#4 mentioned in Section~\ref{problem}, we discuss the overall end-to-end overhead and scalability of our profiling design. We analyzed the base overhead of \name profiling for all possible MPI counters and timers with support of profiling I/O and sending them to the \name using a scientific application on 4,096 processes across 256 nodes.

\begin{figure}[htbp]
    \centering
    \includegraphics[width=0.9\linewidth]{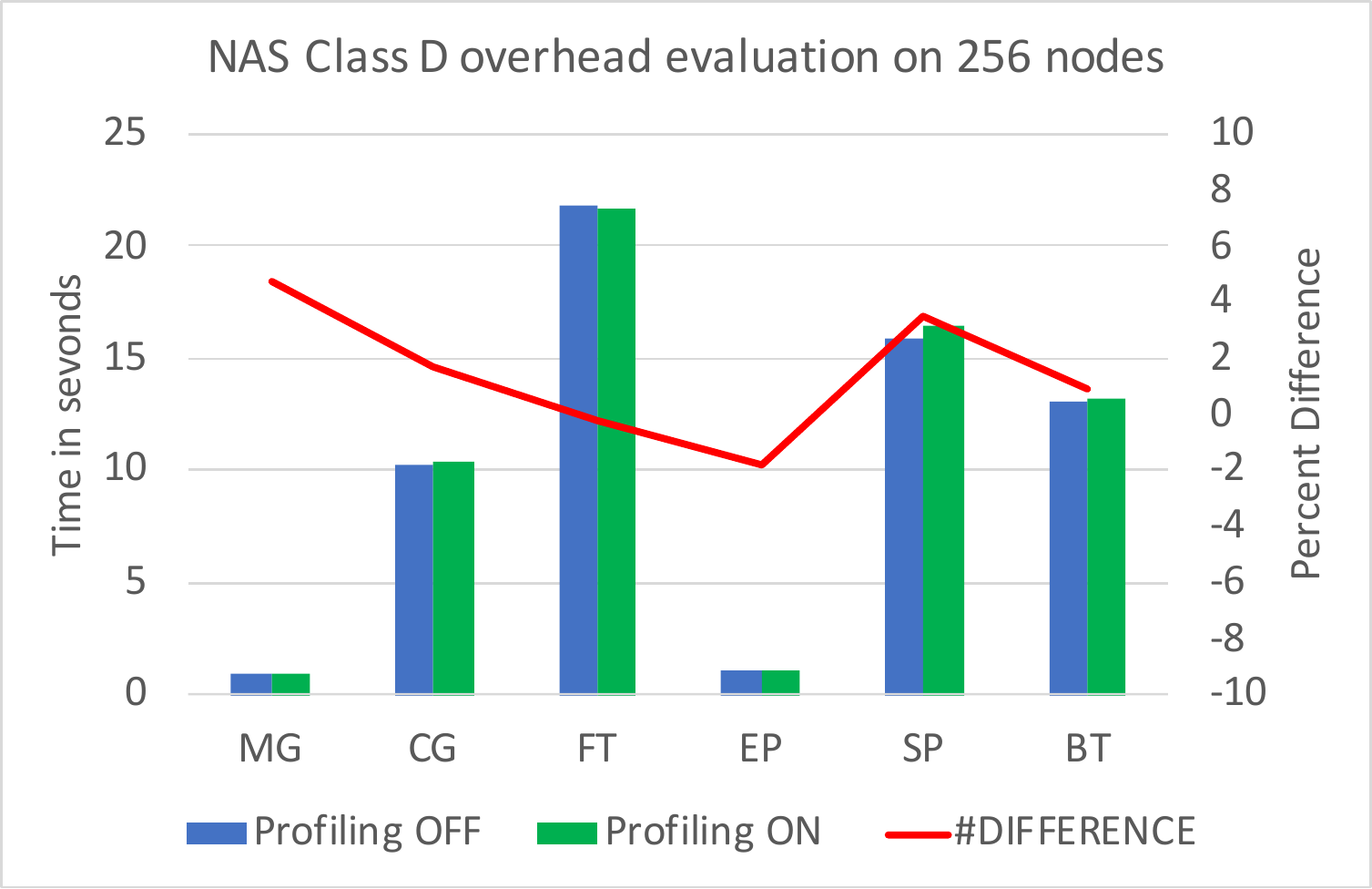}
    \MyCaption{Overhead analysis of \name with class D NAS parallel benchmarks at 4,096 processes across 256 nodes with profiling interval of 5 seconds. There is less than 5\% performance degradation. }
    \label{fig:nas-d}
\end{figure}

Figure \ref{fig:nas-d} compares the
performance of different NAS Class D benchmarks with
and without MPI profiling performance variables (PVAR) with an interval of 5 seconds for 4,096 processes across 256 nodes.
The number are an average of 25 iterations for each benchmark of NAS. 
On a HPC system a 2\% performance jitter is normal. We observe that full-stack profiling with INAM fits well for in-production runs with end-to-end performance overhead at scale.

\MySection{Related work}
\label{sec:related_work}

HPC community is moving toward enabling vast options of profiling tools to measure performance and 
operate HPC systems. As mentioned, Monitoring MPI jobs down to the port counters are
essential for debugging job failures and finding network bottlenecks. In this
section, we give details of existing tools (in literature and those
available as products) besides those already mentioned
in Section~\ref{sec:intro}.
The Texas Advanced Computing Center has developed a tool called \textit{TACC
STATS}~\cite{taccstats}. This tool correlates job and system-level statistics to
create reports. The reports are used to identify issues in the jobs or systems
infrastructure, but, not in real-time. In~\cite{hopsa}, the authors proposed a
suite built on MPI and allows for different metrics for each job and thus may
not provide a full system view of how the jobs are interacting with the fabric.
The authors in~\cite{peruse} describe a library that goes beyond the PMPI interface
to gather MPI states and lower-level network statistics. Paraver~\cite{paraver}
is used to provide visualization of the data. However, the tool could not
visualize / model network activities. Lustre file system records many statistics
that can be extracted and displayed. Some related work
on making these statistics available can be found in~\cite{lustreStats}. 

MonSTer~\cite{monster} suggests using parallel coordinates for measuring node health information like Fan speed, temperature, and intra node metrics but lacks the support for a full system view of the network topology. HiperView \cite{hiperview} extends MonSTer to gather job-level information with node resource utilization and health stats but depends on existing infrastructures to collect metrics which limits the portability. Besides, it does not support online end-to-end profiling. 
As described in Section~\ref{sec:intro}, several tools allow
systems administrators to analyze and inspect high-performance networks such as
Mellanox, FabricIT, BoxFish, Nagios, or Ganglia.  With all of these tools, there
is no integration with the MPI Library, so correlating network traffic to MPI
jobs is a manual endeavor.  On the other hand, there are several tools such as
TAU, HPCToolkit, Intel VTune, IPM, and mpiP focus on the MPI applications
but do not show the network traffic of the MPI jobs.

\MySection{Software Availability, Deployment, and Community Outreach}
\label{sec:software}

The \#2 visualization design (Figures~\ref{fig:topo} and~\ref{fig:lustre-traffic}) with support for Lustre I/O profiling has been released publicly and is available
for free download and use from the project website~\cite{inam}. 
So far this version has had over 850+ downloads from the project site. \name
is currently deployed at the Ohio Supercomputer
Center to monitor multiple HPC clusters. It is also being used at
GeorgiaTech, University of
Utah and at institutions in Poland. We are continually working
with other institutions to deploy the tool on their clusters.
The team actively participates in various outreach activities
including presenting the features and capabilities of the
proposed tool through tutorials held as part of various national
and international conferences/workshops as well as webinars
conducted by organizations like XSEDE~\cite{xsede} and
CaRCC~\cite{CaRCC}.

\MySection{Conclusion}
\label{sec:conclusion}
In this paper, we proposed and implemented new
visualization methods to enable holistic insight for representing the cross-stack metrics of HPC communication stack. Moreover, we propose and implement a low-overhead I/O profiling inside MPI library, collect, send and store the profiling information, and then study the correlation and evaluation of I/O traffic with MPI communication using a cross-stack approach through using INAM. Through experimental evaluations and use cases, we demonstrated novel benefits of our cross-stack communication analysis in real-time to detect bottlenecks and understand communication performance. With new visualization techniques users can: 1) understand their job allocation topology in the system and have more accurate expectations of network performance, jitter, and locality of their nodes. 2) By having the holistic view of communication, users can relate, compare, and understand the interplay of communication of different layers of HPC stack with each other. 3) Benefit from online, stand alone profiling to increase the productivity of domain scientists who are not familiar with gathering, correlating and visualizing the communication information from various HPC stacks. 4) System administrators can identify congested links, nodes, and switches very quickly and take action. We demonstrated low-overhead of less than 5\% for full-stack profiling, gathering and sending communication  metrics for large scale application running on 4,096 processes on 256 nodes. We have scaled the proposed visualizations to large scale ($\approx 2,000~nodes$) HPC clusters. 

As part of future work, we aim to do the following: 1) study other visualization methods to demonstrate jobs statistics in company with network communication visualization. 2) port the
designs as part of \name to other popular tools such as TAU and XDMoD.




\MySection{Acknowledgements}
\label{sec:ack}

We thank Mansa Kedia for her efforts and contributions. 
This research is supported in part by NSF grants \#CNS-1513120,
\#ACI-1450440, \#CCF-1565414, and \#ACI 1664137.

\bibliographystyle{IEEEtran}

\begin{thebibliography}{10}
\providecommand{\url}[1]{#1}
\csname url@samestyle\endcsname
\providecommand{\newblock}{\relax}
\providecommand{\bibinfo}[2]{#2}
\providecommand{\BIBentrySTDinterwordspacing}{\spaceskip=0pt\relax}
\providecommand{\BIBentryALTinterwordstretchfactor}{4}
\providecommand{\BIBentryALTinterwordspacing}{\spaceskip=\fontdimen2\font plus
\BIBentryALTinterwordstretchfactor\fontdimen3\font minus
  \fontdimen4\font\relax}
\providecommand{\BIBforeignlanguage}[2]{{%
\expandafter\ifx\csname l@#1\endcsname\relax
\typeout{** WARNING: IEEEtran.bst: No hyphenation pattern has been}%
\typeout{** loaded for the language `#1'. Using the pattern for}%
\typeout{** the default language instead.}%
\else
\language=\csname l@#1\endcsname
\fi
#2}}
\providecommand{\BIBdecl}{\relax}
\BIBdecl

\bibitem{mpi3}
\BIBentryALTinterwordspacing
{Message Passing Interface Forum}, {Accessed: \today}. [Online]. Available:
  \url{http://www.mpi-forum.org/}
\BIBentrySTDinterwordspacing

\bibitem{ib-multicast}
A.~Mamidala, L.~Chai, H.-W. Jin, and D.~Panda, ``Efficient smp-aware mpi-level
  broadcast over infiniband's hardware multicast,'' in \emph{Proceedings 20th
  IEEE International Parallel Distributed Processing Symposium}, 2006, pp. 8
  pp.--.

\bibitem{sharp}
``{Mellanox Scalable Hierarchical Aggregation and Reduction Protocol
  (SHARP)},''
  \url{https://www.mellanox.com/files/related-docs/prod_acceleration_software/Mellanox_SHARP_SW_API_Guide.pdf}.

\bibitem{gpu-inam}
P.~{Kousha}, B.~{Ramesh}, K.~{Kandadi Suresh}, C.~{Chu}, A.~{Jain},
  N.~{Sarkauskas}, H.~{Subramoni}, and D.~K. {Panda}, ``Designing a profiling
  and visualization tool for scalable and in-depth analysis of high-performance
  gpu clusters,'' in \emph{2019 IEEE 26th International Conference on High
  Performance Computing, Data, and Analytics}, 2019, pp. 93--102.

\bibitem{pearc-inam}
{ P. Kousha, S. D. Kamal Raj, H. Subramoni, D. Panda, H. Na, T. Dockendorf, K.
  Tomko }, ``{ Accelerated Real-time Network Monitoring and Profiling at Scale
  using OSU INAM },'' in \emph{{ Practice and Experience in Advanced Research
  Computing (PEARC 2020) }}, July 2020.

\bibitem{subramoni2016inam}
H.~Subramoni, A.~M. Augustine, M.~Arnold, J.~Perkins, X.~Lu, K.~Hamidouche, and
  D.~K. Panda, ``{INAM 2: InfiniBand Network Analysis and Monitoring with
  MPI},'' in \emph{International Conference on High Performance
  Computing}.\hskip 1em plus 0.5em minus 0.4em\relax Springer, 2016, pp.
  300--320.

\bibitem{tau}
A.~D. Malony and S.~Shende, ``{Performance Technology for Complex Parallel and
  Distributed Systems},'' in \emph{Proc. DAPSYS 2000, G. Kotsis and P. Kacsuk
  (Eds)}, 2000, pp. 37--46.

\bibitem{hpctoolkit}
\BIBentryALTinterwordspacing
{HPCToolkit}, 2019, {Accessed: \today}. [Online]. Available:
  \url{http://hpctoolkit.org/}
\BIBentrySTDinterwordspacing

\bibitem{vtune}
{Intel Corporation}, ``{Intel VTune Amplifier},''
  \url{https://software.intel.com/en-us/intel-vtune-amplifier-xe}.

\bibitem{IPM}
``{Integrated Performance Monitoring (IPM)},''
  {http://ipm-hpc.sourceforge.net/}.

\bibitem{mpip}
``{mpiP: Lightweight, Scalable MPI Profiling},''
  http://\linebreak[0]www.llnl.gov/\linebreak[0]CASC/\linebreak[0]mpip/.

\bibitem{IntelITAC}
{Intel}, ``{Intel Trace Analyzer and Collector},''
  \url{https://software.intel.com/en-us/trace-analyzer}.

\bibitem{arm-map}
{ARM Holdings}, ``{ARM MAP},''
  \url{https://www.arm.com/products/development-tools/server-and-hpc/forge/map}.

\bibitem{hvprof}
A.~A. {Awan}, A.~{Jain}, C.~{Chu}, H.~{Subramoni}, and D.~K. {Panda},
  ``Communication profiling and characterization of deep-learning workloads on
  clusters with high-performance interconnects,'' \emph{IEEE Micro}, vol.~40,
  no.~1, pp. 35--43, 2020.

\bibitem{xdmod}
J.~T. {Palmer}, S.~M. {Gallo}, T.~R. {Furlani}, M.~D. {Jones}, R.~L. {DeLeon},
  J.~P. {White}, N.~{Simakov}, A.~K. {Patra}, J.~{Sperhac}, T.~{Yearke},
  R.~{Rathsam}, M.~{Innus}, C.~D. {Cornelius}, J.~C. {Browne}, W.~L. {Barth},
  and R.~T. {Evans}, ``Open xdmod: A tool for the comprehensive management of
  high-performance computing resources,'' \emph{Computing in Science
  Engineering}, vol.~17, no.~4, pp. 52--62, July 2015.

\bibitem{PCP}
``{Performance Co-Pilot},'' \url{https://pcp.io}.

\bibitem{Prometheus}
``{Prometheus exporter},'' \url{https://github.com/prometheus/node\_exporter}.

\bibitem{fabricit}
{Mellanox Integrated Switch Management Solution},
  \url{http://www.mellanox.com/page/ib\_fabricit\_efm\_management}.

\bibitem{boxfish}
{Lawrence Livermore National Laboratory}, ``{PAVE: Performance Analysis and
  Visualization at Exascale},''
  \url{https://computation.llnl.gov/project/performance-analysis-through-visualization/software.php}.

\bibitem{ldms}
\BIBentryALTinterwordspacing
A.~Agelastos, B.~Allan, J.~Brandt, P.~Cassella, J.~Enos, J.~Fullop, A.~Gentile,
  S.~Monk, N.~Naksinehaboon, J.~Ogden, M.~Rajan, M.~Showerman, J.~Stevenson,
  N.~Taerat, and T.~Tucker, ``{The Lightweight Distributed Metric Service: A
  Scalable Infrastructure for Continuous Monitoring of Large Scale Computing
  Systems and Applications},'' in \emph{SuperComputing 2014}, ser. SC
  '14.\hskip 1em plus 0.5em minus 0.4em\relax Piscataway, NJ, USA: IEEE Press,
  2014, pp. 154--165. [Online]. Available:
  \url{http://dx.doi.org/10.1109/SC.2014.18}
\BIBentrySTDinterwordspacing

\bibitem{collectc}
``{collectc},'' http://collectl.sourceforge.net/.

\bibitem{iml}
``{Integrated Manager For Lustre},''
  https://github.com/whamcloud/integrated-manager-for-lustre.

\bibitem{lmt}
``{Lustre Monitoring Tool},'' https://github.com/LLNL/lmt/wiki.

\bibitem{darshan}
\BIBentryALTinterwordspacing
P.~Carns, K.~Harms, W.~Allcock, C.~Bacon, S.~Lang, R.~Latham, and R.~Ross,
  ``Understanding and improving computational science storage access through
  continuous characterization,'' \emph{ACM Trans. Storage}, vol.~7, no.~3, Oct.
  2011. [Online]. Available: \url{https://doi.org/10.1145/2027066.2027068}
\BIBentrySTDinterwordspacing

\bibitem{TOP500}
``{TOP 500 Supercomputer Sites},'' http://www.top500.org.

\bibitem{mellonox-switch}
\BIBentryALTinterwordspacing
``{Mellanox CS7500 Switch Series},'' {Accessed: \today}. [Online]. Available:
  \url{http://www.mellanox.com/page/products_dyn?product_family=191&mtag=cs7500}
\BIBentrySTDinterwordspacing

\bibitem{parallel-cor}
\BIBentryALTinterwordspacing
``{Parallel Coordinates},'' 2021, {Accessed: \today}. [Online]. Available:
  \url{https://en.wikipedia.org/wiki/Parallel_coordinates}
\BIBentrySTDinterwordspacing

\bibitem{inamdownload}
\BIBentryALTinterwordspacing
``{OSU INAM},'' 2021, {Accessed: \today}. [Online]. Available:
  \url{http://mvapich.cse.ohio-state.edu/tools/osu-inam/}
\BIBentrySTDinterwordspacing

\bibitem{osc}
\BIBentryALTinterwordspacing
``{Ohio Supercomputer Center},'' {Accessed: \today}. [Online]. Available:
  \url{https://www.osc.edu/}
\BIBentrySTDinterwordspacing

\bibitem{taccstats}
{B. Barth, T. Evans and J. McCalpin}, ``Tacc stats,''
  \url{https://www.tacc.utexas.edu/research-development/tacc-projects/tacc-stats}.

\bibitem{hopsa}
{Virtual Institute - High Productivity Supercomputing}, ``{HOPSA: A Holistic
  Performance System Analysis},''
  \url{http://www.vi-hps.org/projects/hopsa/overview}.

\bibitem{peruse}
R.~Keller, G.~Bosilca, G.~Fagg, M.~Resch, and J.~J. Dongarra, ``{Implementation
  and Usage of the PERUSE-Interface in Open MPI},'' in \emph{Proceedings, 13th
  European PVM/MPI Users' Group Meeting}, ser. Lecture Notes in Computer
  Science.\hskip 1em plus 0.5em minus 0.4em\relax Bonn, Germany:
  Springer-Verlag, September 2006.

\bibitem{paraver}
{Barcelona Supercomputing Center}, ``{Paraver},''
  \url{http://www.bsc.es/computer-sciences/performance-tools/paraver}.

\bibitem{lustreStats}
``{Lustre Monitoring and Statistics},''
  http://wiki.lustre.org/Lustre\_Monitoring\_and\_Statistics\_Guide.

\bibitem{monster}
J.~Li, G.~Ali, N.~Nguyen, J.~Hass, A.~Sill, T.~Dang, and Y.~Chen, ``Monster: An
  out-of-the-box monitoring tool for high performance computing systems,'' in
  \emph{2020 IEEE International Conference on Cluster Computing (CLUSTER)},
  2020, pp. 119--129.

\bibitem{hiperview}
N.~Nguyen, T.~Dang, J.~Hass, and Y.~Chen, ``Hiperjobviz: Visualizing resource
  allocations in high-performance computing center via multivariate
  health-status data,'' in \emph{2019 IEEE/ACM Industry/University Joint
  International Workshop on Data-center Automation, Analytics, and Control
  (DAAC)}, 2019, pp. 19--24.

\bibitem{inam}
{OSU InfiniBand Network Analysis and Monitoring Tool},
  \url{http://mvapich.cse.ohio-state.edu/tools/osu-inam/}.

\bibitem{xsede}
``{XSEDE - Extreme Science and Engineering Discovery Environment},''
  \url{https://www.xsede.org/}.

\bibitem{CaRCC}
``{CaRCC - Campus Research Computing Consortium},'' \url{https://carcc.org/}.

\end{thebibliography}


\end{document}